\documentclass[prl,twocolumn,showpacs,floatfix,amsfonts]{revtex4}
\usepackage{graphicx,graphics,color,epsfig,exscale}
\usepackage{bm}
\usepackage{amsmath}
\usepackage{amssymb}

\def\sgn{\text{sgn}}

\def\be{\begin{equation}} 
\def\ee{\end{equation}} 
\def\bea{\begin{eqnarray}} 
\def\eea{\end{eqnarray}}
\begin{document}
\preprint{}
\title{Scaling and Enhanced Symmetry at the Quantum Critical
Point of the Sub-Ohmic Bose-Fermi Kondo Model}
\author{Stefan Kirchner and Qimiao Si}
\affiliation
{Department of Physics \& Astronomy, Rice University, Houston, 
TX 77005, USA}

\begin{abstract}
We consider the finite temperature scaling properties of
a Kondo-destroying 
quantum critical point 
in the Ising-anisotropic 
Bose-Fermi Kondo model (BFKM). A cluster-updating Monte Carlo approach
is used, in order to reliably access a wide temperature range.
The scaling function for the 
two-point spin correlator
is found to have the form dictated by
a boundary conformal field theory, even though the underlying 
Hamiltonian lacks conformal invariance. 
Similar conclusions are reached for all multi-point correlators
of the spin-isotropic BFKM in a dynamical large-N limit.
Our results
suggest that the quantum critical local properties 
of the sub-ohmic BFKM 
are those of an underlying boundary conformal field theory.
\end{abstract}
\maketitle
Quantum criticality is currently being discussed in the contexts of
a wide array of strongly correlated electron systems.
A prototype is provided by a family of 
heavy fermion metals near their antiferromagnetic quantum 
critical point (QCP).
The physical properties of these materials drastically deviate 
from the expectations of the traditional theory of quantum
criticality\cite{Loehneysen.07,Schroder2,Paschen.04,Gegenwart.07}, so much so
that the
question has been raised as to whether and how 
the Kondo effect itself becomes critical at the antiferromagnetic
quantum transition\cite{Si.01,Coleman.01,Senthil.04}.
Through the extended dynamical mean field 
theory, the self-consistent BFKM provides one means to elucidate 
such Kondo-destroying quantum criticality\cite{Si.01}.
The sub-Ohmic BFKM is also the appropriate low-energy model for
single-electron transistors attached to ferromagnetic
leads\cite{Kirchner.05a}.
One clue\cite{Zhu.04,Vojta.05,Glossop.05}
about the nature of the quantum criticality in
the sub-Ohmic BFKM is the failure
of the standard description\cite{Hertz.76}
in terms of fluctuations of the classical order 
parameter in elevated dimensions.
Nonetheless, a proper field theory
for the QCP
is not yet available. To address this pressing open issue,
it is important to identify the symmetry of the QCP.

In this letter,
we study the finite temperature scaling properties of the BFKM
in some detail. We have been motivated by general considerations of 
a boundary conformal field theory\cite{Affleck.91,Cardy.84}.
The latter arises in many quantum impurity problems whose 
bulk system in the continuum limit is conformally invariant.
At zero temperature,
the s-wave component of the bulk degrees of freedom can be
thought of as living on a half-plane, which is 
composed of the imaginary time $(\tau)$ dimension and 
the radial spatial ($r$) dimension; the 
quantum impurity is located at the boundary of the
half-plane\cite{Affleck.91}.
At finite temperature ($T$), the extent along the imaginary
time direction becomes finite, of length $\beta=1/T$,
and periodic boundary condition (along $\tau$) turns it
into a half-cylinder of circumference $\beta$.
A conformal mapping between the half-plane and the 
half-cylinder\cite{Affleck.91,Ginsparg.88},
say $z=\tan(\pi w /\beta)$, 
can then be used to obtain finite-temperature correlators from
their zero-temperature counterparts. The result is the well
known scaling form\cite{Tsvelik,Ginsparg.88}
\begin{equation}
<\Phi(\tau,T)\Phi(0,T)>\,=\,C \Big(\frac{\pi/\beta}{\sin(\pi\tau/\beta)}
\Big)^{2\Delta},
\label{scaling-cft-general}
\end{equation}
where $\Delta$ is the scaling dimension of $\Phi$, a conformal
primary field, and $C$ a constant.

We present results here which show that the scaling functions of
the two-point spin correlators of the 
sub-Ohmic BFKM have the form dictated by Eq.~(\ref{scaling-cft-general}).
Similar conclusions are drawn for multi-spin correlators of 
the model in a large-N limit.
These results are
surprising,
since the sub-Ohmic nature 
[Eq.~(\ref{EQ:sub-Ohmic}), with $\epsilon >0$]
of the bosonic spectrum implies that the bulk component of the 
Hamiltonian itself lacks conformal invariance.
The results imply that the symmetry is enhanced at the boundary 
QCP of the BFKM,
in such a way that the local properties
are those of an underlying boundary conformal field theory (CFT).

Our focus will be the Ising-anisotropic spin-$1/2$ BFKM. 
In order to address the finite
temperature scaling properties, it is important to access 
a wide temperature range with sufficiently high accuracy.
Here, we develop a cluster-updating Monte Carlo method,
and show that it can reliably reach temperatures as low
as $10^{-4}~T_K^0$,
where 
$T_K^0$ is the
Kondo scale of the fermion-only Kondo problem.
The wide temperature range covered distinguishes this method
from existing ones for Kondo-type
systems\cite{Hirsch.86,Grempel.99,Grempel.03}.

{\it BFKM with Ising Anisotropy:~}
In a BFKM, a quantum spin is simultaneously coupled to
a fermionic bath and a bosonic one. 
For the Ising-anisotropic case, the Hamiltonian is
\begin{eqnarray}
{\cal H}_{\text{\small bfkm}} =&& J_K ~{\bf S}
\cdot {\bf s}_c + \sum_{p\sigma}
E_{p}~c_{p\sigma}^{\dagger}~ c_{p\sigma}
\nonumber\\ &&
+ \; \tilde{g} \sum_{p} S^z \left( \phi_{p} + \phi_{-p}^{\;\dagger}
\right) + \sum_{p} w_{p}\,\phi_{p}^{\;\dagger} {\phi}_{p}\; ,
\label{EQ:H-imp}
\end{eqnarray}
where  ${\bf S}$ is a spin-$1/2$ local moment,
$c_{p\sigma}^{\dagger}$ describes a fermionic bath
with a constant density 
of states, $\sum_{p} \delta (\omega - E_{p}) = N_0$,
and $\phi_{p}^{\;\dagger}$ a bosonic bath
whose spectrum is sub-Ohmic ($\epsilon>0$):
\begin{equation}
\sum_p [\delta(\omega-\omega_p)- \delta(\omega+\omega_p)] \sim
|\omega|^{1-\epsilon} \sgn(\omega) .
\label{EQ:sub-Ohmic}
\end{equation}
We adopt bosonization and a canonical 
transformation\cite{Grempel.03} 
to map ${\cal H}_{\text{\small bfkm}}$ to
\begin{eqnarray}
{\cal H}_{\text{\small bfkm}}'
 =&&
\Gamma S^x +
\Gamma_z S^z_{ } s^z_c
+ {\cal H}_0^{}
\nonumber\\ &&
+ \;  \tilde{g}\sum_{p} S^z \left( \phi_{p} + \phi_{-p}^{\;\dagger}
\right) + \sum_{p} w_{p}\,\phi_{p}^{\;\dagger} {\phi}_{p},
\label{EQ:H-imp'}
\end{eqnarray}
where ${\cal H}_0$ and $s^z_c$ describe the (bosonized) fermionic bath
and local conduction electron spin.
The quantities $\Gamma$ and ${\Gamma_z}$ are, respectively,
determined by the spin-flip and longitudinal components of the Kondo
coupling.
Integrating out both the fermionic and bosonic baths, we
arrive at the partition function $Z_{\text{\small bfkm}}' \sim
{\text{Tr}}~ {\rm exp} [-{\cal S}_{\text{\small imp}}']$ with
\begin{eqnarray}
{\cal S}_{\text{\tiny imp}}'\!\!&=&\!\! \int_0^{\beta} d \tau 
[ \Gamma S^x(\tau) -{1 \over 2} \int_0^{\beta} d \tau'\!
S^z(\tau) S^z(\tau')
\nonumber\\
&&\times (\chi_0^{-1}(\tau-\tau') - {\cal K}_c(\tau-\tau') ) ],
\label{eff-action}
\end{eqnarray}
 where the trace is
taken over spin degrees of freedom.
$\chi_0^{-1} = \tilde{g}^2 \sum _{p} G_{\phi,0}$,
and ${\cal K}_c (i\omega_n) = \kappa_c |\omega_n| $
($\kappa_{c}\sim \Gamma_{z}^{2}N_{0}^{2}$)
came from integrating out the fermionic bath.
Trotter decomposing the effective action and re-expressing the leading
order (in $1/L$, $L$ being the number of time slices)
through the transfer matrix for a one-dimensional Ising model\cite{Blume.70},
one finally obtains,
\begin{equation}
{\mathcal{Z}}
\sim {\text{Tr}}~\exp\big[\sum_{i} K_{\text {NN}} S_{i}
S_{i+1} + \sum_{i,j} K_{\text {LR}}(i-j)S_{i}S_{j} \big] .
\label{classicalIsing}
\end{equation} 
The mapping procedure is essentially equivalent to 
what was done for the pure Kondo model\cite{Anderson.69}.
The effective action 
at inverse temperature $\beta$
is equivalent to that of a one-dimensional chain of $L$ Ising
spins, with a periodic boundary condition.
The nearest neighbor interaction is $K_{\text
  {NN}}=-\ln(\tau_0\Gamma/2)/2$, where $\tau_0=\beta/L$;
it 
is singular in the limit
$\tau_0 \rightarrow 0$.
$K_{\text
 { LR}}(i-j)$ is the sum of two  ferromagnetic  
long-ranged interactions proportional to
$1/|i-j|^2$ and $1/|i-j|^{2-\epsilon}$; it results from discretizing 
$(\chi_0^{-1}(\tau-\tau') - {\cal K}_c(\tau-\tau') )$:
\begin{equation}
K_{\mbox{\tiny LR}}(|i-j|)= \frac{\tau_0^2}{4}\Big[
\frac{2\alpha  (\pi/\beta)^2}{\sin(\frac{\pi \tau_0
    |i-j|}{\beta})^2}+ \frac{ g
  (\pi/\tilde{\beta})^{2-\epsilon}}{\sin(\frac{\pi \tau_0
    |i-j|}{\tilde{\beta}})^{2-\epsilon}} \Big].
\end{equation}
The coupling constant $\alpha$ is related to the electron scattering
phase shift. We choose $\alpha=1/2$, so that $g=0$ corresponds to the
Toulouse limit of the Kondo problem.
For the most part, the parameter $\tilde{\beta}\gg \beta$ is taken to be 
20000$\tau_0$.
(In the cases we have checked, we found identical 
results when the second term in the brackets is simply replaced by
$g/\tau^{2-\epsilon}$.)
The thermodynamic limit  has to be taken
such that a finite $T_K^0$ is preserved, in order to have the
temperature region of interest, $T<T_K^0$; in the limit
$\tau_0 \rightarrow 0$ at finite $\beta$ we approach
the high-temperature fixed point. For the numerical values 
chosen in the manuscript we are always able to focus on the
scaling properties in the low temperature range of $T<<T_K^0$.

We 
study this model using
a cluster-updating 
Monte Carlo (MC) scheme. The long-range
nature of the
interaction is most conveniently incorporated using 
the method of Ref.~\cite{Luijten.95}.
We specifically use a Wolff algorithm\cite{Wolff.88}.
The
improved estimator for the spin-spin correlation function
implies that the susceptibility in Matsubara frequency is
given by
\begin{equation}
\label{improved_estimator}
\chi(i\omega_n)\,=\, <|\sum_{j\varepsilon {\mathcal{C}}} e^{i
  \omega_n \tau_j}|^2/n_{\mathcal{C}}>/TL,
\end{equation}
where the sum runs over all spins in the cluster, $<>$ indicates the
average over all Monte Carlo runs and $n_{\mathcal{C}}$
is the number of spins in a given cluster. The
susceptibility as a function of imaginary time $\tau$ 
is given by
$
\label{tau_measure}
\chi(\tau_n)\,=\, \frac{1}{4}\sum_{S_m^z,S^z_{m+n}\varepsilon {\mathcal{C}}}
<S^z(\tau_m+\tau_n)S^z(\tau_m)>
$, where $\tau_n=n\tau_0$.
We measure $\chi(i\omega_n)$
and  $\chi(\tau_n)$ directly.

This method allows us to reach considerably lower temperatures than
approaches using local updates\cite{Grempel.99,Grempel.03}.
We typically build 2000 MC clusters as a warm-up and about $N_{run}=10^6$ 
MC clusters at  high
temperatures and increase the number of clusters built to about 50000 warm-ups 
and about $N_{run}=10^{10}$  clusters at $\beta=512$ for all $\tau_0$. 
While every cluster built contributes to  
$\chi(\tau= 0)$  only the
subset of clusters with spins separated by $\tau\ge\beta/2$ will
contribute to $\chi(\tau \approx\beta/2)$. One might therefore expect
that variance and autocorrelation effects strongly depend on
$\tau$, 
but this turned out not 
to be the case. 
For an error estimate,
we performed a binning analysis of our data in order to obtain the integrated
autocorrelation time $\tau_{int}$ and  variance
\footnote[1]{The integrated
  autocorrelation time of $\chi$ is defined as $\tau_{int}= \frac{1}{2}\sum_i
  (<\chi_k \chi_{k+i}>-<\chi>^2)/(<\chi^2>-<\chi>^2)$, where $\chi_k$
  and $\chi_{k+1}$ are successive measurements of $\chi$.}.
The relative error of our
results is $(\Delta \chi)/\chi \approx 10^{-2}$ and below (depending on
$\tau$ and $\beta$) and the
integrated autocorrelation time is $\tau_{int}<100\ll N_{run}$
 for all $\tau$ and $\beta$.\\
For concreteness, we will now present the results for $\epsilon=0.4$.

\begin{figure}[t!]
\includegraphics[width=0.5\textwidth]{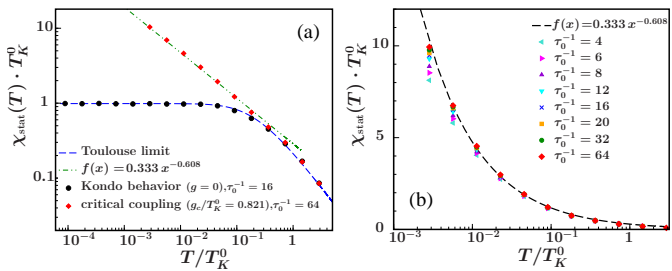}
\caption{(a) Static local spin susceptibility in the Kondo case ($g=0$,
$\tau_0=1/16$, black circles) and for the critical coupling ($\epsilon=0.4,
\Gamma=0.75, g_c=0.821T_K^0, \tau_0=1/64$, red  diamonds).
The dashed blue line is the fit to the Toulose limit
(see e.g. Ref.~\cite{Guinea.85}),
and the dashed dotted
line a fit to the critical behavior. Defining
$T_K^0 \equiv 1/\chi_{\mbox{stat}}(T=0)$
we obtain 
$T_K^0 \tau_0\approx 0.688$; 
(b) Static susceptibility at the critical coupling, for 
various
values of the cutoff parameter, $\tau_0$.
}
\label{static-chi}
\end{figure}

Consider first the Kondo limit ($g=0$). 
In Fig.~\ref{static-chi}(a)
we show the static spin susceptibility
versus temperature.
It correctly captures the Pauli behavior at temperatures
below $T_K^0$. Because we have placed the Kondo couplings at the
Toulouse point, we can compare our results with the exact expression:
Fig.~\ref{static-chi}(a) demonstrates the agreement for more than
4 decades of temperature!
In addition, 
our results for the dynamical
susceptibility\cite{Kirchner_upl.07}
are consistent with the standard expectations for the
Kondo problem, including the asymptotic long time 
(low frequency) Fermi-liquid power-law behavior and
the exact limit at short time,
$\chi (\tau \rightarrow \tau_0) \rightarrow 1/4$.

\begin{figure}[t!]
\includegraphics[width=0.36\textwidth]{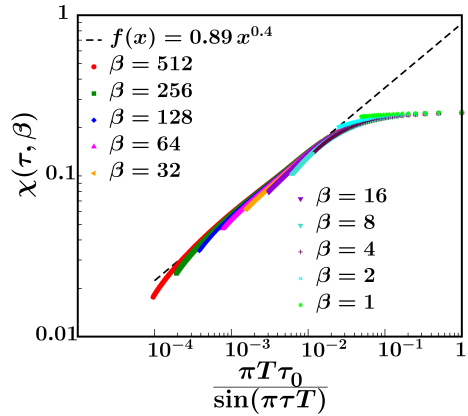}
\caption{Scaling of the local spin susceptibility 
(at the QCP),
which is plotted as a function of 
$\pi T \tau_0 /\sin(\pi \tau T)$.
The parameters are $\epsilon=0.4,\Gamma=0.75, g=0.821T_K^0\approx
g_c, \tau_0=1/64$. Note that 
$\pi T \tau_0 /\sin(\pi \tau T)$ 
becomes small (order $\pi \tau_0/\beta$)
as $\tau$ approaches the long-time limit, $\tau \rightarrow \beta/2$.
The power-law collapse occurs over two decades of the parameter
$\pi T \tau_0 /\sin(\pi \tau T)$. The deviation at the low-left corner
is attributed to finite-size effects.
}
\label{chi_tau}
\end{figure}

Fig.~\ref{static-chi}(a) also shows the static susceptibility at the 
QCP,
$g_{c}/T_K^0=0.821$. 
We find $\chi_{stat}(T,g_c)
\sim 1/T^{0.608}$, for over two decades of temperature.
Since $\chi_{}(\omega,g_c)\sim 1/\omega^{1-\epsilon}$ is
expected\cite{Zhu.02,Zarand.02},
the temperature exponent is, within about 1\% accuracy,
the same as the frequency exponent; this is consistent
with the NRG result\cite{Glossop.05}.
The dependence of the critical susceptibility on the cutoff
parameter $\tau_0$ is illustrated in Fig.~\ref{static-chi}(b).
Compared to the Kondo case (not shown), this dependence is 
stronger. However, within the measured temperature range,
the result does not change significantly for the smallest
three $\tau_0$ values.
We now turn to the $\tau$-dependence of the dynamical spin susceptibility,
$\chi(\tau,T)$, near the quantum critical coupling $g \approx g_c$.
The scaling plot, Fig.~\ref{chi_tau}, demonstrates that $\chi(\tau,T)$
is a function of $\pi T/\sin (\pi \tau  T)$ only. 
In the long-time limit (lower-left corner), the dependence is a simple
power-law for over two decades 
of $\pi T/\sin(\pi \tau T)$. 
We therefore reach one of our key results,
namely
\begin{equation}
  \chi_{crit}(\tau,T)\,=\, \Phi\big(\frac{\pi\tau_0 T}{\sin (\pi \tau
  T)}\big)\,\stackrel{T\ll T_K^{0}}{\longrightarrow}\, 
c \cdot \big(\frac{\pi\tau_0 T}{\sin (\pi \tau
  T)}\big)^{\epsilon},
\label{scalingform}
\end{equation}
for $\tau^{-1}_{} \ll T_K^0$; here $c$ is a constant ($\approx 0.89$).
In other words, in the long-time (low-energy) limit, the two-point spin
correlator has precisely the form dictated by a boundary CFT,
inspite of the lack of conformal invariance in the 
Hamiltonian. This scaling form implies $\omega/T$-scaling.

We now turn to complementary results on 
the multi-point correlators, which 
are provided by the large-N limit of a spin-isotropic BFKM.

{\it Spin-isotropic BFKM in a large-N limit:~}
The limit is taken for the 
Hamiltonian of the 
 SU(N)$\times$S(M) BFKM,
\begin{eqnarray}
{\cal H}_{\text{MBFK}} &=&
({J_K}/{N})
\sum_{\alpha}{\bf S}
\cdot {\bf s}_{\alpha}
+ \sum_{p,\alpha,\sigma} E_{p}~c_{p \alpha
\sigma}^{\dagger} c_{p \alpha \sigma}
\nonumber\\
&+&
({g}/{\sqrt{N}})
{\bf S} \cdot
{\bf \Phi}
+ \sum_{p}
w_{p}\,{\bf \Phi}_{p}^{\;\dagger}\cdot {\bf \Phi}_{p},
\label{H-MBFK}
\end{eqnarray}
where the spin and channel indices
are $\sigma = 1, \ldots, N$
and
$\alpha=1, \ldots, M$,
respectively,
and ${\bf \Phi} \equiv \sum_p ( {\bf \Phi}_{p} +
{\bf \Phi}_{-p}^{\;\dagger} )$ contains $N^2-1$
components.  This dynamical large-N limit\cite{Parcollet.98,Cox.93}
is expressed in terms of pseudo-fermions $f_{\sigma}$ and a bosonic
decoupling field $B_{\alpha}$, where 
$S_{\sigma,\sigma^{\prime}}=f^{\dagger}_{\sigma}f^{}_{\sigma^{\prime}}
-\delta_{\sigma,\sigma^{\prime}}Q/N$,
where $Q$ is related to the chosen irreducible representation 
of SU(N)\cite{Parcollet.98}.
The large-$N$ equations are
\begin{eqnarray}
\Sigma_B(\tau) &=& - {\cal G}_{0}(\tau) G_f(-\tau);
 \nonumber \\
\Sigma_f(\tau)&=& \kappa {\cal G}_{0}(\tau) G_B(\tau) + g^2
G_f(\tau){\cal G}_{\Phi}(\tau); \nonumber \\
G_B^{-1}( i\omega_n) &=&
1/{J_K} - \Sigma_B( i\omega_n); \nonumber \\
G_f^{-1}(i\omega_n)& =& i\omega_n - \lambda -
\Sigma_f(i\omega_n);
\label{NCA}
\end{eqnarray}
together with a constraint $G_f(\tau\rightarrow 0^{-})=Q/N$.
Here, $\kappa=M/N$, $\lambda$ is a Lagrangian multiplier,
${\cal G}_0 = - \langle T_{\tau} c_{\sigma\alpha}(\tau)
c_{\sigma\alpha}^{\dagger}(0) \rangle _0$,
and ${\cal G}_{\Phi} =  \langle  T_{\tau} \Phi(\tau)
\Phi^{\dagger}(0) \rangle _0$. Note that,
when $g=0$, the Kondo Hamiltonian contains a
conformally-invariant bulk
and the corresponding correlation functions 
naturally have the form of a boundary
CFT\cite{Parcollet.98,Cox.93}.
Here we address what happens at the QCP of the 
model with finite $g$~\cite{Zhu.04}, for which the bulk
lacks conformal invariance.

Consider first the zero-temperature case.
The quantum critical properties of the model 
have been determined
in Ref.~\cite{Zhu.04}. 
At $g=g_c$, both the pseudo-fermion propagator $G_f(\tau)$
and the auxiliary boson propagator $G_B(\tau)$ are critical;
their leading terms 
are $G_f(\tau) = A/|\tau|^{\epsilon/2} \sgn(\tau)$ and 
$G_B(\tau) = B/|\tau|^{1-\epsilon/2}$,
respectively.
Here, we observe that the local two-spin correlator,
as well as all the local higher-multiple-spin correlators,
factorize in terms of $G_f(\tau)$ according to Wick's theorem.
This immediately implies that the scaling functions for 
all these correlators have the form 
of a boundary CFT.
The dynamical spin susceptibility,
{\it e.g.}, is
\begin{eqnarray}
\chi (\tau) \equiv 
\langle  T_{\tau} S_{\sigma \neq \sigma'} (\tau)
S_{\sigma' \sigma} (0) 
\rangle 
 = - G_f (\tau) G_f (-\tau) ,
\label{chi-tau}
\end{eqnarray}
whose leading behavior is $1/\tau^\epsilon$.
Likewise, the three-point correlator is
$\sim 1/|\tau_{12} \tau_{13} \tau_{23}|^{\epsilon/2}$,
and the four-point correlator 
$\sum_{i<j}\tau_{ij}^{-\epsilon/3} F(x)$;
here the cross-ratio $x=\tau_{12}\tau_{34}/\tau_{13}\tau_{24}$,
and $\tau_{ij}\equiv \tau_i-\tau_j$.
All these are consistent with the general form of a 
(boundary) CFT\cite{Ginsparg.88}. 

At finite temperatures we solve 
equations (\ref{NCA})
on real frequencies.
The numerical parameters are as in Ref.~\cite{Zhu.04}:
We choose
$\kappa =1/2$,
$Q/N=1/2$, and
 $N_0(\omega) = (1/\pi){\rm exp}(-\omega^2/\pi)$
for the conduction electron density of states.
The nominal bare Kondo scale is
$T_K^0 N_0(0)\equiv {\rm exp}(-1/N_0(0)J_K) \approx 0.02$,
for fixed $J_K N_0(0)=0.8/\pi$.
The bosonic bath spectral function
$\sum_{p} \delta(\omega-w_{p}) \sim \omega^{1-\epsilon}_{}$
is cut off smoothly
at $2 \omega N_0(0)\approx 0.05$.
The imaginary time
correlation functions are then obtained from
\begin{equation}
\Phi(\tau)\,=\, -\eta \int_{-\infty}^{\infty}
d\omega\,
\frac{\exp(-\tau \omega)}{\exp(-\beta\omega)-\eta}
{\text{Im}} (\Phi(\omega+i0^+)) ,
\end{equation}
for $0<\tau \leq\beta$. Here, 
$\eta=\pm$ for bosonic/fermionic $\Phi$.

\begin{figure}[t!]
\includegraphics[width=0.5\textwidth]{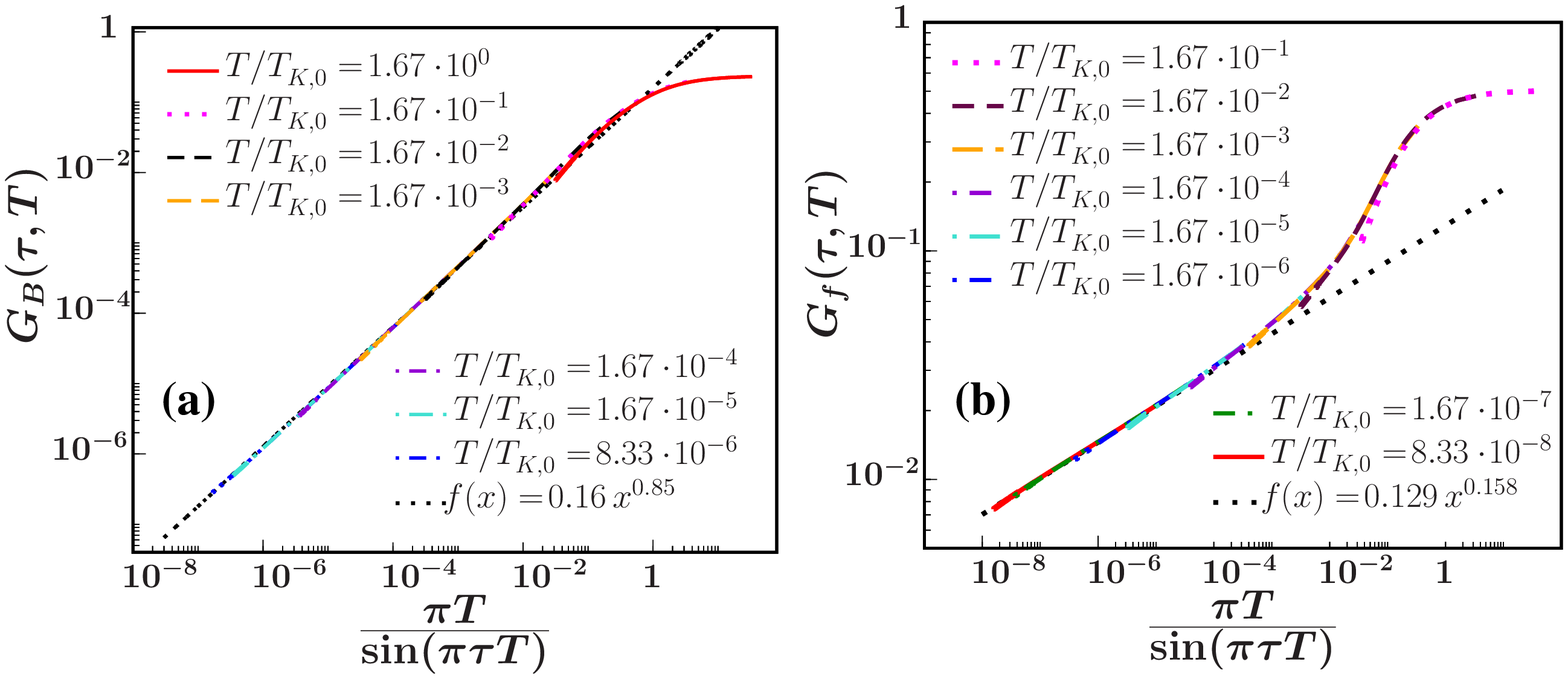}
\caption{Scaling of the propagators for the auxiliary boson,
$G_B(\tau)$ [panel (a)] and for the pseudo-fermion,  
$G_f(\tau)$ [panel (b)],
for $\epsilon=0.3$ and the numerical parameters specified
in the main text, at the critical coupling $g_c=25.5 T_K^0$.}
\label{largeN}
\end{figure}

Figs.~\ref{largeN}(a) and \ref{largeN}(b) show the scaling functions
for $\epsilon=0.3$. For over four (five) 
decades of $\pi T/\sin(\pi \tau T)$,
$G_{f}(\tau)$ [$G_{B}(\tau)$] satisfies the 
conformal form  of
Eq.~(\ref{scaling-cft-general}). (The critical exponents
are compatible with the aforementioned analytical results,
although correction to scaling is somewhat larger in 
$G_{f}$ than in $G_{B}$.) 
Because of their Wick factorizability in terms of 
$G_{f}(\tau)$,
all the finite-temperature local multi-spin correlation
functions will 
assume 
the form of a boundary CFT.

Symmetry enhancement at a fixed point is known
to happen in other contexts.
Moreover, in the case of ordinary (classical) critical points,
it is already known that scale invariance is generically
accompanied by conformal invariance\cite{Ginsparg.88}.
What is nontrivial here is that the continuum limit of the 
bulk part of the Hamiltonian lacks conformal invariance.
Our results suggest that, even in this case, the boundary
correlators of the boundary QCP can be described in terms of those
of an effective model with conformal invariance.

In summary, we have 
studied the finite-temperature  quantum critical properties of the
BFKM.
Our results suggest that the quantum critical
point of the BFKM has an enhanced symmetry. This insight is expected
to be important for the understanding of the underlying field theory
of this Kondo-destroying quantum critical point.

We 
thank C.~J.~Bolech, 
H.~G.~Evertz and, especially,
A.~W.~W.~Ludwig for 
useful
discussions.
This work has been supported in part by
NSF Grant No. DMR-0706625, the Robert A. Welch Foundation,
the W. M. Keck Foundation,
the Rice Computational Research Cluster
funded by NSF
and a partnership between Rice University, AMD and Cray,
and (for S.K.) DOE Grant No. DE-FG-02-06ER46308.

\end{document}